\documentclass[a4paper,11pt]{article}
\usepackage{tikz}
\usepackage[compat=1.1.0]{tikz-feynman}
\usepackage{jheppub} 
\usepackage{verbatim}
\usepackage[T1]{fontenc} 
\usepackage{amsmath}
\usepackage{braket}
\usepackage{float}
\usepackage{slashed}
\usepackage{graphicx} 
\usepackage{tabularx}
\usepackage{booktabs} 
\usepackage{braket}
\usepackage{mathtools}
\usepackage{caption,subcaption}
\usepackage{axodraw4j}
\usepackage{pstricks}
\usepackage{color}
\usepackage{braket}
\usepackage{cleveref}
\usepackage{aligned-overset}
\usepackage{multirow}
\usepackage{breqn}
\usepackage{listings}
\usepackage{autobreak}

\title{\boldmath Static Potential of the Standard Model and Spontaneously Broken Theories}

\author[a]{B. Assi,}
\author[a]{B.A. Kniehl}

\affiliation[a]{II. Institut f\"{u}r Theoretische Physik, Universit\"{a}t Hamburg, 22761 Hamburg, Germany}

\abstract{We consider the static potential in theories exhibiting spontaneous symmetry breaking. We use our findings to calculate the static potential of the Standard Model at one-loop order. We do so in both the Wilson loop and scattering amplitude approaches and discuss the limitations of the Wilson loop approach. As the field content of the SM is extensive, analogous results to ours in a large set of models is now achievable by varying the appropriate couplings and group theory factors.  }
\begin{document} 
\maketitle
\flushbottom
\newpage
\section{Introduction}
\label{sec:intro}
The static potential is a crucial quantity for quantum field theories as it represents the interaction energy of a pair of heavy particles. The potential allows one to study the fundamental properties of a given theory in the non-relativistic (NR) limit. The most well-known static potentials are the Coulomb potential in QED and the non-Abelian analogue of QCD. The QCD static potential for a pair of heavy quarks is known to N3LO \cite{smirnov2010three} and valuable in the study of NR bound systems, such as quarkonia. It is of importance in many areas, such as quark mass definitions and quark pair-production at threshold \cite{czarnecki1998two}. The static potential has also been studied in the context of BSM theories for predicted heavy particles such as in MSSM and $\mathcal{N}=4$ SYM \cite{maldacena1998wilson,arkani2009theory,beneke2020wino}.

We begin by focusing on the QCD static potential, which is of leading importance to heavy quark theory due to the dominance of the strong coupling in the SM. The original idea of describing a bound state of heavy coloured objects, in analogy to the hugely successful Hydrogen atom, was proposed by Susskind in his 1970 Les Houches lecture \cite{susskind1977weak}. In order to demonstrate asymptotic freedom in Yang-Mills theory, he computed the one-loop pole terms using a Wilson loop formula for the potential, and in the process re-derived the first coefficient of the renormalization group Beta function. More recently, the two and three-loop corrections were discovered and turned out to be numerically significant triggering several investigations in further contexts \cite{schroder1999static,smirnov2010three,anzai2010static,peter1997static,brambilla2010three}.

It is expected that the potential consists of two terms: a Coulomb-like short distance term which is perturbatively calculable; and a long-distance term responsible for the phenomenon of quark confinement \cite{bander1981theories}. Thus, a perturbative analysis will not provide the full potential and may not hold the key to gaining a deeper understanding of confinement; the short-distance part may still be employed as a starting point for constructing potential models which have been vastly successful in the description of quarkonia \cite{fischler1992quark}. Moreover, it provides an excellent description for very heavy systems such as $t\Bar{t}$ to high accuracy. The potential in perturbative calculations is comparable with results from numerical calculations in lattice gauge theory.

The Wilson loop approach first employed by Susskind continues to be used to this day due to its computational simplicity. In this approach the static potential in coordinate space, $V(\boldsymbol{r})$, is defined in terms of a Wilson loop, $W(\boldsymbol{r},T)$, with small but finite spatial extension, $\boldsymbol{r}$, and temporal extension, $T\rightarrow\infty$ \cite{schroder1999static}. In this limit $W(\boldsymbol{r},T)\sim \exp{(-iT V(\boldsymbol{r}))}$ and the potential in momentum space, $V(\boldsymbol{q})$, is simply its Fourier transform. However, there has always been discussion about whether the Wilson-loop formula is well-defined due to possible infrared divergences at higher orders \cite{appelquist1978static}. On the other hand, the scattering amplitude approach yields identical results and involves a computation of the on-shell quark anti-quark scattering amplitude and directly yields the momentum-space static potential in the non-relativistic, $\boldsymbol{q}\rightarrow 0$, limit.

In this paper, we consider extending the static potential to theories that exhibit spontaneous symmetry breaking (SSB); in particular, we take on the case of the standard model (SM). The only case of a static potential in the context of a theory with SSB was in the seminal result by Maldacena for heavy W-bosons in $\mathcal{N}=4$ SYM \cite{maldacena1998wilson}. Working off of their result, we attempted an analogous procedure to obtain a SM potential; however, limitations became apparent, which we discuss in detail. Whence, instead, we employed the scattering amplitude approach, which provided us with the full SM static potential to one-loop order. Furthermore, due to the richness of the SM in its field content, it becomes simple to compute static potentials in other theories by a replacement of the appropriate couplings and group theory factors. We then demonstrate applications of our result to beyond QCD corrections in heavy quark effective theories and threshold mass schemes \cite{beneke1998quark,hoang1999b,brambilla2018relations}, in particular the popular 1S and potential subtracted (PS) mass definitions. We found that, as is to be expected from previous results on EW corrections to short-distance heavy-quark mass definitions \cite{kniehl1996dependence}, the EW regime contributes at the same order as NNLO pure QCD contributions to the static potential. Therefore, it stands to reason that they must be incorporated into high precision threshold calculations \cite{beneke2015next}.
\section{Wilson Loop Appraoch}
\label{sec:WilsonLoop}
Let us consider a system with an arbitrary field, $\psi(x)$, defined by an action, $I[\psi]=\int d^d x \mathcal{L}[\psi]$ in the presence of external sources, $J(x)$. One can express the ground state energy of this system in QFT as follows \cite{fischler1992quark},
\begin{equation}
  -\lim_{T\rightarrow \infty}\frac{1}{T}\log{\frac{\int\mathcal{D}\lbrace \psi\rbrace\exp{\left\lbrace-\int d^dx\left[\mathcal{L}(\psi)+J(x)\psi(x)\right]\right\rbrace}}{\int\mathcal{D}\lbrace \psi\rbrace\exp{\left\lbrace-\int d^dx\mathcal{L} \right\rbrace}}},
  \label{eqn:gen}
\end{equation}
and outside the time interval $(-\frac{1}{2}T,\frac{1}{2}T)$, the sources have been switched off. This formula in perturbation theory has been proven exactly for the case of a linear local coupling between the field and external source \cite{symanzik1970small}. We generalise if we assume that in all cases the vacuum-to-vacuum transition amplitude is given by,
\begin{equation}
    \braket{0^+|0^-}_J=\frac{\int\mathcal{D}\lbrace \psi\rbrace\exp{\left\lbrace -S_0[\psi]+S_{int}[\psi,J]\right\rbrace}}{\int\mathcal{D}\lbrace \psi\rbrace\exp{\left\lbrace-S_0[\psi]\right\rbrace}},
\end{equation}
then inserting a complete set of energy eigen-states we may write,
\begin{equation}
    \braket{0^+|0^-}_J=\bra{0}e^{-HT}\ket{0}=\sum_n\bra{0}e^{-HT}\ket{n}\braket{n|0}=\sum_n\left|\braket{0|n}\right|^2e^{-E_nT},
\end{equation}
where the smallest energy eigenvalue, $E_0$, corresponds to the ground state, and thus, in the limit $T\rightarrow\infty$ it dominates the sum. Whence taking the logarithm and dividing by $(-T)$ provides one with the ground state energy, \eqref{eqn:gen}, as is well known \cite{bander1981theories}.

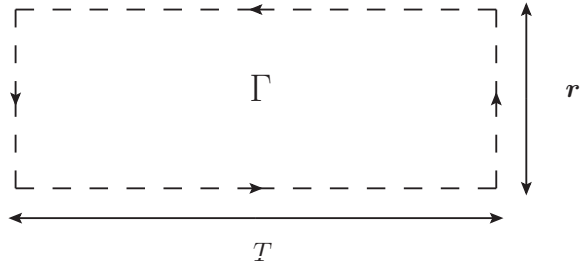
\begin{figure}[tbp]
    \centering
\begin{center}
    \scalebox{0.7}{
\fcolorbox{white}{white}{
  \begin{picture}(334,134) (236,-76)
    \SetWidth{1.0}
    \SetColor{Black}
    \Line[dash,dashsize=10,arrow,arrowpos=0.5,arrowlength=5,arrowwidth=2,arrowinset=0.2](240,-42)(496,-42)
    \Line[dash,dashsize=10,arrow,arrowpos=0.5,arrowlength=5,arrowwidth=2,arrowinset=0.2](496,-42)(496,54)
    \Line[dash,dashsize=10,arrow,arrowpos=0.5,arrowlength=5,arrowwidth=2,arrowinset=0.2](496,54)(240,54)
    \Line[dash,dashsize=10,arrow,arrowpos=0.5,arrowlength=5,arrowwidth=2,arrowinset=0.2](240,54)(240,-42)
    \Line[arrow,arrowpos=1,arrowlength=5,arrowwidth=2,arrowinset=0.2](368,-58)(496,-58)
    \Line[arrow,arrowpos=1,arrowlength=5,arrowwidth=2,arrowinset=0.2](368,-58)(240,-58)
    \Line[arrow,arrowpos=1,arrowlength=5,arrowwidth=2,arrowinset=0.2](512,6)(512,54)
    \Line[arrow,arrowpos=1,arrowlength=5,arrowwidth=2,arrowinset=0.2](512,6)(512,-42)
    \Text(368,-81)[lb]{\Large{\Black{$T$}}}
    \Text(535,7)[lb]{\Large{\Black{$\boldsymbol{r}$}}}
    \Text(366,5)[lb]{\huge{\Black{$\Gamma$}}}
  \end{picture}}
}
\end{center}
    \caption{Rectangular Wilson loop to be integrated over.}
    \label{fig:Wloop}
\end{figure}

We may now be more specific in our discussion and take a gauge field theory, QED for instance, where the energy we calculate corresponds to a system of photons interacting with two point-like static electric charges (with identical magnitude but opposite sign),
\begin{equation}
    -\lim_{T\rightarrow \infty}\frac{1}{T}\log{\frac{\int\mathcal{D}\lbrace \psi\rbrace\exp{\left\lbrace-\int d^dx\left[-\frac{1}{4}F_{\mu\nu}^2+\frac{1}{2\eta}(\partial_{\mu}A_{\mu})^2+J_{\mu}(x)A^{\mu}(x)\right]\right\rbrace}}{\int\mathcal{D}\lbrace \psi\rbrace\exp{\left\lbrace-\int d^dx -\frac{1}{4}F_{\mu\nu}^2+\frac{1}{2\eta}(\partial_{\mu}A_{\mu})^2 \right\rbrace}}},
    \label{eqn:abgs}
\end{equation}
such that,
\begin{equation}
    J_{\mu}(x)=g\delta_{\mu 0}[\delta(\boldsymbol{x})-\delta(\boldsymbol{x}-\boldsymbol{r})]\theta(T^2/4-x_0^2).
\end{equation}
We may then re-write the numerator of Eq. \eqref{eqn:abgs} as the following expectation value,
\begin{equation}
    \left\langle \mathcal{T}\exp{\left\lbrace g\int(A_0(t,\boldsymbol{r})-A_0(t,\boldsymbol{0}))dt\right\rbrace}\right\rangle,
\end{equation}
where $\mathcal{T}$ stands for time ordering. This Green's function is manifestly gauge invariant, which one can see by considering the gauge invariant operator, $ \mathcal{P} \exp{\left[\oint_{\Gamma}g A_{\mu} dx^{\mu}\right]}$, where $\mathcal{P}$ denotes path ordering and $\Gamma$ is the rectangular loop of spatial and time extent, $\boldsymbol{r}$ and $T$, respectively, as in figure \ref{fig:Wloop}. Now, in the limit, $T\rightarrow \infty$, the spatial components, $\lbrace\boldsymbol{A}(\xi,\frac{1}{2}T), \boldsymbol{A}(\xi,-\frac{1}{2}T)\rbrace$, reduce to pure gauge terms as the field strength tensor, $F_{\mu\nu}=0$ at infinity and thus is gauge equivalent to $\boldsymbol{A}=0$. Therefore, the operator, $\mathcal{T}\exp{\left\lbrace g\int(A_0(\boldsymbol{r},t)-A_0(\boldsymbol{0},t))dt\right\rbrace}$, is gauge invariant and so is the ground state energy (or static potential), which is equal to,
\begin{equation}
   V(\boldsymbol{r})= -\lim_{T\rightarrow \infty}\frac{1}{T}\log{\frac{ \left\langle \mathcal{T}\exp{\left\lbrace g\int(A_0(t,\boldsymbol{r})-A_0(t,\boldsymbol{0}))dt\right\rbrace}\right\rangle}{\langle\boldsymbol{1} \rangle}}.
\end{equation}
This approach has been employed in QED where only the LO term contributes to all orders \cite{dyson1952divergence}, and the non-Abelian case of QCD has been studied to three loop orders \cite{smirnov2010three}. We begin by re-evaluating the QCD case and then extending this approach to theories with SSB as in the case of the SM and beyond. 
\subsection{QED/QCD}
\label{ssec:WLQCD}
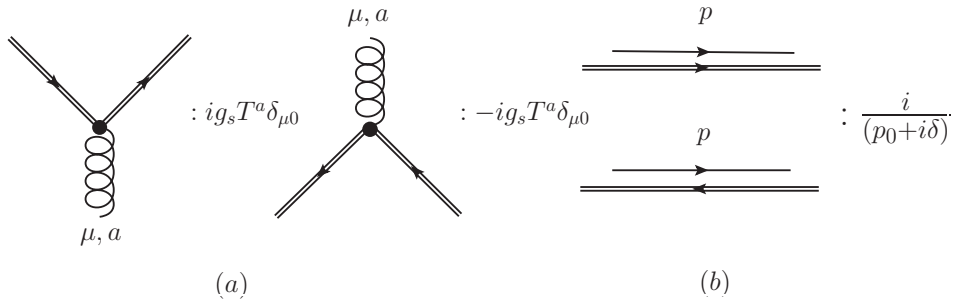
\begin{figure}[tbp]
\centering
\begin{center}
\scalebox{0.7}{
\fcolorbox{white}{white}{
  \begin{picture}(501,173) (94,5)
    \SetWidth{1.0}
    \SetColor{Black}
    \Line[arrow,arrowpos=0.5,arrowlength=5,arrowwidth=2,arrowinset=0.2,double,sep=2](96,140)(144,92)
    \Line[arrow,arrowpos=0.5,arrowlength=5,arrowwidth=2,arrowinset=0.2,double,sep=2](144,92)(192,141)
    \Gluon(144,92)(144,44){7.5}{4}
    \Gluon(288,140)(288,93){7.5}{4}
    \Line[arrow,arrowpos=0.5,arrowlength=5,arrowwidth=2,arrowinset=0.2,double,sep=2](287,92)(238,44)
    \Line[arrow,arrowpos=0.5,arrowlength=5,arrowwidth=2,arrowinset=0.2,double,sep=2](336,45)(289,92)
    \Vertex(144,92){4}
    \Vertex(288,91){4.123}
    \Line[arrow,arrowpos=0.5,arrowlength=5,arrowwidth=2,arrowinset=0.2,double,sep=2](401,124)(528,124)
    \Line[arrow,arrowpos=0.5,arrowlength=5,arrowwidth=2,arrowinset=0.2,double,sep=2](527,59)(400,59)
    \Line[arrow,arrowpos=0.5,arrowlength=5,arrowwidth=2,arrowinset=0.2](417,69)(512,69)
    \Line[arrow,arrowpos=0.5,arrowlength=5,arrowwidth=2,arrowinset=0.2](417,133)(514,132)
    \Text(206,0)[lb]{\Large{\Black{$(a)$}}}
    \Text(465,1)[lb]{\Large{\Black{$(b)$}}}
    \Text(134,29)[lb]{\Large{\Black{$\mu,a$}}}
    \Text(277,147)[lb]{\Large{\Black{$\mu,a$}}}
    \Text(465,148)[lb]{\Large{\Black{$p$}}}
    \Text(464,83)[lb]{\Large{\Black{$p$}}}
    \Text(193,92)[lb]{\Large{\Black{$:ig_sT^a\delta_{\mu 0}$}}}
    \Text(337,92)[lb]{\Large{\Black{$:-ig_sT^a\delta_{\mu 0}$}}}
    \Text(540,83)[lb]{\huge{\Black{$:\frac{i}{(p_0+i\delta)}$}}}
  \end{picture}}
}
\end{center}
\caption{Feynman rules for (anti-)source propagator and (anti-)source-gluon vertices.}
\label{fig:sRules}
\end{figure}

\begin{figure}[bp]
\centering
\begin{center}
\scalebox{0.6}{
\fcolorbox{white}{white}{
  \begin{picture}(256,136) (83,-27)
    \SetWidth{1.0}
    \SetColor{Black}
    \Line[arrow,arrowpos=0.5,arrowlength=5,arrowwidth=2,arrowinset=0.2,double,sep=2](192,88)(288,88)
    \Gluon(192,88)(192,-24){7.5}{9}
    \Line[arrow,arrowpos=0.5,arrowlength=5,arrowwidth=2,arrowinset=0.2,double,sep=2](96,88)(192,88)
    \Line[arrow,arrowpos=0.5,arrowlength=5,arrowwidth=2,arrowinset=0.2,double,sep=2](192,-24)(96,-24)
    \Line[arrow,arrowpos=0.5,arrowlength=5,arrowwidth=2,arrowinset=0.2,double,sep=2](288,-24)(192,-24)
    \Text(55,88)[lb]{\Large{\Black{$(E,\boldsymbol{p})$}}}
    \Text(294,88)[lb]{\Large{\Black{$(E,\boldsymbol{p'})$}}}
    \Text(210,27)[lb]{\Large{\Black{$(0,|\boldsymbol{p'}-\boldsymbol{p}|)$}}}
    \Text(45,-24)[lb]{\Large{\Black{$(E,-\boldsymbol{p})$}}}
    \Text(294,-24)[lb]{\Large{\Black{$(E,-\boldsymbol{p'})$}}}
  \end{picture}}
}
\end{center}
\caption{Tree diagram for the QCD static potential.}
\label{fig:SPtree}
\end{figure}
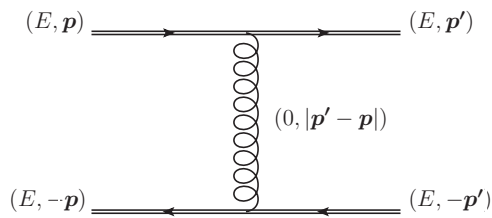

Let us begin by considering the potential in QCD \cite{schroder1999static}, which corresponds to the interaction energy of an infinitely massive $Q\overline{Q}$ pair separated by a fixed distance, $\boldsymbol{r}$, interacting by exchanging virtual gluons. Using the definition,
\begin{subequations}
    \begin{align}
        V(\boldsymbol{r})&=-\lim_{T\rightarrow \infty}\frac{1}{T}\log{\left\langle \mathcal{W}[\Gamma]\right\rangle} \\ &
        =-\lim_{T\rightarrow \infty}\frac{1}{T}\log{\left\langle\overline{\text{tr}}\mathcal{P}\exp{\left(i g \oint_{\Gamma} d^4x J^{\mu}A_{\mu}\right)}\right\rangle}
        ,
    \end{align}
\end{subequations}
such that $\mathcal{W}[\Gamma]$ denotes the Wilson loop, $\mathcal{P}$ denotes path ordering, $\overline{\text{tr}}$ denotes the normalised color trace, $\overline{\text{tr}}(...)\equiv \text{tr}(...)/\text{tr}{(\boldsymbol{1})}$, and the gauge potential, $A_{\mu}(x)=T^a_{ij}A^a_{\mu}(x)$. $\Gamma$ is the rectangular Wilson loop as shown in figure \ref{fig:Wloop}, and 
\begin{equation}
    \langle\mathcal{O}(A)\rangle\equiv\frac{\int\mathcal{D}A\exp{(-S)}\mathcal{O}(A)}{\int\mathcal{D}A\exp{(-S)}}.
\end{equation}
The desired properties of static colour charge is dictated by,
\begin{equation}
    J^{\mu}(x)=v^{\mu}[\delta(\boldsymbol{x})-\delta(\boldsymbol{x}-\boldsymbol{r})]\theta(T^2/4-x_0^2),
\end{equation}
such that $v^{\mu}\equiv\delta^{\mu 0}$. After Fourier transforming to momentum space, we get the Feynman rules for our potential \cite{schroder1999static}. The QCD Feynman rules remain unaltered aside for those illustrated in figure \ref{fig:sRules}.
To illustrate computing the Wilson loop, consider the tree amplitude illustrated in figure \ref{fig:SPtree}, where $|\boldsymbol{p}|=|\boldsymbol{p'}|$ and $E=\sqrt{m^2+\boldsymbol{p}^2}$. 
From this tree diagram one obtains the following amplitude,
\begin{equation}
    i\mathcal{M}=i\frac{4\pi\alpha_s}{|\boldsymbol{k}-\boldsymbol{k}'|^2}T^a_{c_1'c_1}T^a_{c_2'c_2}\equiv i\frac{4\pi\alpha_s}{\boldsymbol{q}^2}\left(\delta_{c_1c_2}\delta_{c_1'c_2'}-\frac{1}{N_c}\delta_{c_1c_1'}\delta_{c_2c_2'}\right),
\end{equation}
where $c_{1(2)}$ and $c_{1(2)}'$ denote colours of initial and final states, respectively. 
The colour singlet case, $c_1=c_2$ and $c_{1(2)}'$ summed over gives $V_s(\boldsymbol{q})=-\frac{C_F\alpha_s}{\boldsymbol{q}^2}$, versus the colour octet case,  $(c_1=c_1',c_2=c_2')$, with no sum, gives $V_o(\boldsymbol{q})=\frac{1}{2C_A}\frac{\alpha_s}{\boldsymbol{q}^2}$, where $C_A=N_c$ and $C_F=(N_c^2-1)/(2N_c)$. 

\begin{figure}[bp]
\centering
\begin{center}
\scalebox{0.5}{
\fcolorbox{white}{white}{
  \begin{picture}(739,361) (31,-27)
    \SetWidth{1.0}
    \SetColor{Black}
    \Line[arrow,arrowpos=0.5,arrowlength=5,arrowwidth=2,arrowinset=0.2,double,sep=2](32,329)(112,329)
    \Line[arrow,arrowpos=0.5,arrowlength=5,arrowwidth=2,arrowinset=0.2,double,sep=2](112,329)(192,329)
    \Line[arrow,arrowpos=0.5,arrowlength=5,arrowwidth=2,arrowinset=0.2,double,sep=2](192,201)(112,201)
    \Line[arrow,arrowpos=0.5,arrowlength=5,arrowwidth=2,arrowinset=0.2,double,sep=2](112,201)(32,201)
    \Line[arrow,arrowpos=0.5,arrowlength=5,arrowwidth=2,arrowinset=0.2,double,sep=2](224,329)(304,329)
    \Line[arrow,arrowpos=0.5,arrowlength=5,arrowwidth=2,arrowinset=0.2,double,sep=2](304,329)(384,329)
    \Line[arrow,arrowpos=0.5,arrowlength=5,arrowwidth=2,arrowinset=0.2,double,sep=2](304,201)(224,201)
    \Line[arrow,arrowpos=0.5,arrowlength=5,arrowwidth=2,arrowinset=0.2,double,sep=2](384,201)(304,201)
    \Line[arrow,arrowpos=0.5,arrowlength=5,arrowwidth=2,arrowinset=0.2,double,sep=2](416,329)(496,329)
    \Line[arrow,arrowpos=0.5,arrowlength=5,arrowwidth=2,arrowinset=0.2,double,sep=2](496,329)(576,329)
    \Line[arrow,arrowpos=0.5,arrowlength=5,arrowwidth=2,arrowinset=0.2,double,sep=2](496,201)(416,201)
    \Line[arrow,arrowpos=0.5,arrowlength=5,arrowwidth=2,arrowinset=0.2,double,sep=2](576,201)(496,201)
    \Line[arrow,arrowpos=0.5,arrowlength=5,arrowwidth=2,arrowinset=0.2,double,sep=2](32,121)(112,121)
    \Line[arrow,arrowpos=0.5,arrowlength=5,arrowwidth=2,arrowinset=0.2,double,sep=2](112,121)(192,121)
    \Line[arrow,arrowpos=0.5,arrowlength=5,arrowwidth=2,arrowinset=0.2,double,sep=2](112,-7)(32,-7)
    \Line[arrow,arrowpos=0.5,arrowlength=5,arrowwidth=2,arrowinset=0.2,double,sep=2](192,-7)(112,-7)
    \Line[arrow,arrowpos=0.5,arrowlength=5,arrowwidth=2,arrowinset=0.2,double,sep=2](416,121)(496,121)
    \Line[arrow,arrowpos=0.5,arrowlength=5,arrowwidth=2,arrowinset=0.2,double,sep=2](496,121)(576,121)
    \Line[arrow,arrowpos=0.5,arrowlength=5,arrowwidth=2,arrowinset=0.2,double,sep=2](496,-7)(416,-7)
    \Line[arrow,arrowpos=0.5,arrowlength=5,arrowwidth=2,arrowinset=0.2,double,sep=2](576,-7)(496,-7)
    \Line[arrow,arrowpos=0.5,arrowlength=5,arrowwidth=2,arrowinset=0.2,double,sep=2](608,121)(688,121)
    \Line[arrow,arrowpos=0.5,arrowlength=5,arrowwidth=2,arrowinset=0.2,double,sep=2](688,121)(768,121)
    \Line[arrow,arrowpos=0.5,arrowlength=5,arrowwidth=2,arrowinset=0.2,double,sep=2](688,-7)(608,-7)
    \Line[arrow,arrowpos=0.5,arrowlength=5,arrowwidth=2,arrowinset=0.2,double,sep=2](768,-7)(688,-7)
    \SetWidth{3.0}
    \Text(201,265)[lb]{\Large{\Black{$;$}}}
    \Text(305,176)[lb]{\Large{\Black{$(a)$}}}
    \Text(497,177)[lb]{\Large{\Black{$(b)$}}}
    \Text(689,176)[lb]{\Large{\Black{$(c)$}}}
    \Text(112,-32)[lb]{\Large{\Black{$(d)$}}}
    \Text(305,-32)[lb]{\Large{\Black{$(e)$}}}
    \Text(496,-31)[lb]{\Large{\Black{$(f)$}}}
    \Text(689,-32)[lb]{\Large{\Black{$(g)$}}}
    \SetWidth{1.0}
    \Gluon(113,329)(113,201){7.5}{10}
    \Gluon(252,329)(252,201){7.5}{10}
    \Gluon(355,328)(355,201){7.5}{10}
    \Gluon(443,330)(554,201){7.5}{14}
    \Gluon(443,201)(556,329){7.5}{14}
    \Line[arrow,arrowpos=0.5,arrowlength=5,arrowwidth=2,arrowinset=0.2,double,sep=2](609,328)(689,328)
    \Line[arrow,arrowpos=0.5,arrowlength=5,arrowwidth=2,arrowinset=0.2,double,sep=2](689,328)(769,328)
    \Line[arrow,arrowpos=0.5,arrowlength=5,arrowwidth=2,arrowinset=0.2,double,sep=2](689,200)(609,200)
    \Line[arrow,arrowpos=0.5,arrowlength=5,arrowwidth=2,arrowinset=0.2,double,sep=2](769,200)(689,200)
    \SetWidth{1.4}
    \Arc[arrow,arrowpos=0.5,arrowlength=8.16,arrowwidth=3.264,arrowinset=0.2](689,264)(32,180,540)
    \SetWidth{1.0}
    \Line[arrow,arrowpos=0.5,arrowlength=5,arrowwidth=2,arrowinset=0.2,double,sep=2](224,121)(304,121)
    \Line[arrow,arrowpos=0.5,arrowlength=5,arrowwidth=2,arrowinset=0.2,double,sep=2](304,121)(384,121)
    \Line[arrow,arrowpos=0.5,arrowlength=5,arrowwidth=2,arrowinset=0.2,double,sep=2](304,-7)(224,-7)
    \Line[arrow,arrowpos=0.5,arrowlength=5,arrowwidth=2,arrowinset=0.2,double,sep=2](384,-7)(304,-7)
    \Gluon(689,296)(689,328){7.5}{3}
    \Gluon(689,231)(689,199){7.5}{3}
    \Gluon(113,121)(113,88){7.5}{3}
    \Gluon(113,24)(113,-7){7.5}{2}
    \Gluon(306,56)(248,120){7.5}{7}
    \Gluon(306,56)(366,120){7.5}{7}
    \Gluon(306,55)(306,-8){7.5}{5}
    \GluonArc[clock](499.13,74.312)(77.717,143.077,36.006){7.5}{12}
    \Gluon(497,121)(498,-7){7.5}{10}
    \Gluon(690,121)(689,57){7.5}{5}
    \Gluon(690,-7)(690,56){7.5}{5}
    \GluonArc(713,56)(25.02,-2,358){7.5}{13}
    \GluonArc(113,56)(30.806,-13,347){7.5}{15}
  \end{picture}}
}
\end{center}
\caption{Diagrams that contribute to the QCD static potential at one-loop. The arrowed circle represents light quark and ghost loops. }
\label{fig:SPQCD}
\end{figure}
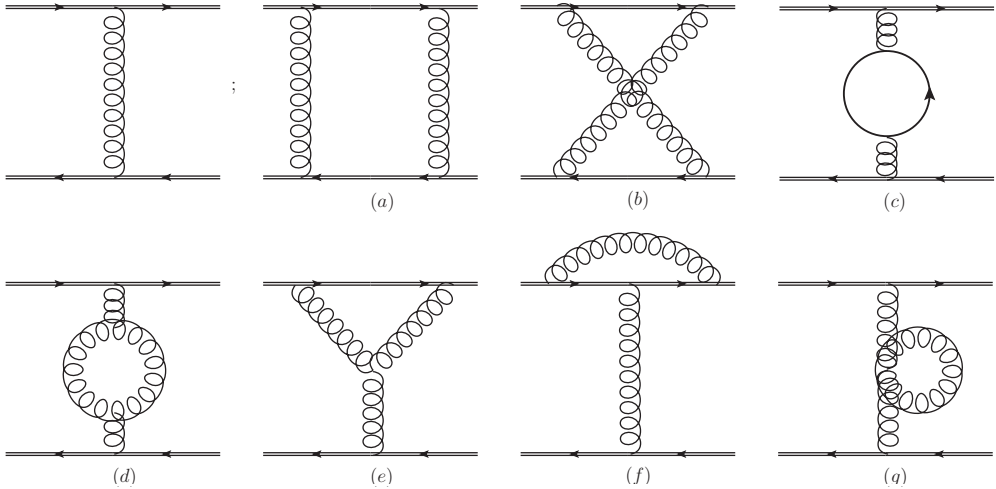

At one-loop in QCD we have amplitudes illustrated in figure \ref{fig:SPQCD}, such that post-reduction, in Feynman Gauge, only (a-d) are non-zero since the remaining ones are scaleless. We re-calculate the one-loop result in the $\overline{\text{MS}}$ scheme and find the well-known quantity \cite{fischler1992quark}. A suggestive way of writing the final result in momentum space is,
\begin{subequations}
\begin{align}
    & V(\boldsymbol{q}^2)=-C_F\frac{4\pi\alpha_V(\boldsymbol{q}^2)}{\boldsymbol{q}^2}, \\ &
    \alpha_V(\boldsymbol{q}^2)=\alpha_s(\mu^2)\sum_{n=0}^{\infty}\Tilde{a}_n(\mu^2/\boldsymbol{q}^2)\left( \frac{\alpha_s(\mu^2)}{4\pi} \right)^n = \alpha_s(\boldsymbol{q}^2)\sum_{n=0}^{\infty}a_n\left( \frac{\alpha_s(\boldsymbol{q}^2)}{4\pi} \right)^n,
\end{align}
\end{subequations}
where $a_0=\Tilde{a}_0=1$ and,
\begin{equation}
    a_1=\frac{31}{9}C_A-\frac{20}{9}T_fn_f \quad,\quad \Tilde{a}_1=a_1+\beta_0\log{\frac{\mu^2}{\boldsymbol{q}^2}}.
\end{equation}
The coupling, $\alpha_s$, denotes the strong coupling in the $\overline{\text{MS}}$ scheme and $\alpha_V$ represents the effective coupling constant which incorporates all radiative corrections into its definition. This provides a new scheme, the V-scheme \cite{barry2000analytical,kataev2015fourth}, which defines the strong coupling in terms of a potential. With the QCD result at hand, one expects to be able to extend this approach to the SM and other theories exhibiting SSB. 


\subsection{$\mathcal{N}=4$ SYM}
\label{ssec:WLN4}
The only case of a static potential being calculated for a spontaneously broken theory with a Higgs-like field in 3+1 dimensional $\mathcal{N}=4$ SYM, which has been done with the Wilson loop approach \cite{maldacena1998wilson}. The static potential in this theory is given by,
\begin{equation}
    V(\boldsymbol{r})=-\lim_{T\rightarrow\infty}\frac{1}{T}\ln{\left\langle \overline{\text{tr}}\mathcal{P}\exp{i\left(\oint_{\Gamma}ds A^{\mu}\dot{x}_{\mu}+\Phi_i\theta_i|\dot{x}|\right)}\right\rangle},
\end{equation}
where $A_{\mu}=A_\mu^aT^a$ is the gauge field, $\Phi_{i=1,\ldots,6}$ are the six scalar fields in this theory and the rectangular Wilson loop to be integrated over is as previously shown in figure \ref{fig:Wloop}. We will summarise the derivation of this potential. Consider the theory with spontaneous breaking of $\text{U(N+1)} \rightarrow \text{U(N)}\times \text{U(1)}$ by giving some expectation value, $\langle\Phi\rangle^i=v^i$, to a Higgs-like field. Then the massive $W$-bosons of $\mathcal{N}=4$ SYM have a mass proportional to $|v|$ and transform in the fundamental representation of $\text{U(N)}$. So in the limit $|v|\rightarrow \infty$ they provide the very massive quarks necessary to compute Wilson loops in the $\text{U(N)}$ theory. The physics of interest is for energy scales much lower than $|v|$ so that the $\text{U(N)}$ theory is effectively decoupled from the $\text{U(1)}$ theory.

Consider the equation of motion for the massive $W$-boson, extracting the leading time dependence as $W=e^{-i|v|t}\Tilde{W}$, we get an equation of motion from the Lagrangian for $\Tilde{W}$, which to leading order in large $|v|$ reads,
\begin{equation}
    \label{eqn:speom}
    (\partial_0-iA_0-i\theta^i\Phi^i)\Tilde{W}=0,
\end{equation}
where we have defined $\theta^i\equiv v^i/|v|$. Notice that $A_0$ and $\Phi^i$ are matrices in the adjoint representation of $\text{U(N)}$. This implies that if we consider this massive $W$-boson describing a closed loop, $\Gamma$, its interaction with the $\text{U(N)}$ gauge field will lead to the insertion of the Wilson loop operator,
\begin{equation}
    \mathcal{W}(\Gamma)=\overline{\text{tr}}\mathcal{P}\exp{i\left(\oint_{\Gamma}ds A^{\mu}\dot{x}_{\mu}+\Phi_i\theta_i|\dot{x}|\right)}.
\end{equation}
This operator is determined by the contour $\Gamma$ (parametrised by $x^{\mu}(s)$) as well as a function, $\theta^i(s)$, which is a unit six-vector (i.e. $\theta^i\theta^i=1$). From this Wilson loop one obtains the static potential by taking the expectation value and limit,
\begin{equation}
    V(\boldsymbol{r})=-\lim_{T\rightarrow \infty}\frac{1}{T}\log{\left\langle \mathcal{W}[\Gamma]\right\rangle}.
\end{equation}
This potential has been evaluated in detail and limits have been mapped to classical D-string solutions \cite{maldacena1998wilson}. More recently this same potential has been computed in the weak-coupling limit to NLO using effective theory methods inspired by pNRQCD \cite{pineda2008static}.

Following this case, we may apply the same procedure to the equation of motion of heavy quarks in the standard model obtained from the SM Lagrangian. The leading time dependence is exhibited analogously for the heavy quarks, $Q=e^{-im_Qt}\Tilde{Q}$, such that $m_Q=\sqrt{2}Yv$, where $Y$ is the quark Yukawa coupling and $v$ is the Higgs vacuum expectation value. The analogous limit we can then consider, $v\sim m_Q, m_{W,Z,H} \gg |\boldsymbol{q}|$ where $|\boldsymbol{q}|$ is the transform momentum between the static sources. 

\subsection{Standard Model}
\label{ssec:WLSM}
Since the Wislon loop approach is technically simpler, we apply it to heavy quarks in the standard model, inspired by the $\mathcal{N}=4$ SYM derivation.
For illustration purposes we will omit couplings to the $W/Z$ and Goldstone bosons as we will see later that they can not be taken into account in this approach.  The quark field, $\psi(x)$, then has equation of motion,
\begin{equation}
    \left(i\slashed{D}-m_Q-\sqrt{2}YH(x) \right)\psi(x)=0,
    \label{eqn:eomwl}
\end{equation}
where $\slashed{D}=\gamma_{\mu}(\partial^{\mu}-i\tilde{A}^{\mu}(x))$ and $\tilde{A}^{\mu}=g_sA^{\mu,a}_gT^a_{ij}-eA_e^{\mu}$ represents the massless gauge field couplings, $H(x)$ corresponds to a gauge singlet scalar field that models Higgs exchange in the SM, the quark Yukawa coupling is given by $Y$, and $m_Q=\sqrt{2}Yv$ where $v$ is the Higgs vacuum expectation value \cite{schwartz2014quantum}. Expanding \eqref{eqn:eomwl} component-wise gives,
\begin{equation}
    \left(i\gamma^{0}\partial_{0}-i\boldsymbol{\gamma}\cdot\boldsymbol{\partial}-\gamma^0\Tilde{A}_0(x)+{\boldsymbol{\gamma}}\cdot\boldsymbol{\Tilde{A}}(x)-\sqrt{2}Y(H(x)+v)\right)\psi(x)=0,
      \label{eqn:eomw2}
\end{equation}
which we can further reduce by solving the Schr\"{o}dinger equation for the heavy quark field, providing us with the leading time dependence, 
\begin{equation}
\psi=e^{-im_Qt}\Tilde{\psi}=e^{(-i\sqrt{2}Yvt)}\Tilde{\psi},
\end{equation}
and plugging this back into \eqref{eqn:eomw2} gives,
\begin{equation}
 \left((\gamma_0-1)\sqrt{2}Yv -i\boldsymbol{\gamma}\cdot\boldsymbol{\partial}-\gamma^0\Tilde{A}_0(x)+{\boldsymbol{\gamma}}\cdot\boldsymbol{\Tilde{A}}(x)-\sqrt{2}YH(x)\right)\Tilde{\psi}(x)=0,
\end{equation}
Taking the limit $v\rightarrow \infty$ of this expression, we attain the bi-spinor constraint, $(1-\gamma_0)\Tilde{\psi}=0$ which forces zero first component, $\Tilde{\psi}=(0,\chi)$. Therefore, all terms acted on by the matrices, $\lbrace {\boldsymbol{\gamma}},\gamma_5\rbrace$, do not contribute, restricting the naive inclusion of $W/Z$ and Goldstone couplings as they are chiral and flavour changing.  We are left with the equation of motion,
\begin{equation}
    \left(i\partial_{t}-\Tilde{A}_0-\sqrt{2}YH\right)\chi(x)=0,
\end{equation}
which may be re-written in the following way,
\begin{equation}
    i\partial_t\chi=\hat{M}\chi \Rightarrow \chi=\hat{U}(t)\chi_0.
\end{equation}
Thus, if we consider this heavy quark describing a closed loop, $\Gamma$, its interaction with the Higgs and gauge fields leads to the insertion of the Wilson loop operator,
\begin{equation}
    \mathcal{W}(\Gamma)=\overline{\text{tr}}\mathcal{P}\exp{i\left(\oint_{\Gamma}d\tau \Tilde{A}^{\mu}(x)\dot{x}_{\mu}+\sqrt{2} YH(x)|\dot{x}|\right)},
\end{equation}
and the potential is then,
\begin{equation}
    V(\boldsymbol{r})=-\lim_{T\rightarrow \infty}\frac{1}{T}\log{\left\langle \mathcal{W}[\Gamma]\right\rangle},
\end{equation}
 in the large $v$ limit. We may also consider the large Yukawa coupling limit. In this case the simplification of the bi-spinor to one large component ceases to occur. Instead, spatio-temporal mixing in spinor components happens resulting in the loss of gauge invariance. Otherwise, in the large $v$ limit we have a potential allowing for interactions between the static source and $\lbrace g,\gamma,H\rbrace$ bosons. Evaluating this potential in momentum space to one-loop order in the $\overline{\text{MS}}$ scheme gives the following extension to the pure QCD result,
\begin{equation}
     V(\boldsymbol{q})= V^{QCD}(\boldsymbol{q})-\frac{4\pi\alpha}{\boldsymbol{q}^2}\left[\frac{4}{9}+\frac{\alpha_s}{4\pi}b_1+\frac{\alpha}{4\pi}b_2\right]
\end{equation}
such that,
\begin{subequations}
\begin{align}
    &b_1=\frac{\tilde{Y} C_F}{\sin{\theta_W}^2}\log{\frac{m_H^2}{\mu^2}} \\&
    b_2=\frac{80}{243}n_f\log{\frac{\mu^2}{\boldsymbol{q}^2}}-\frac{400}{729}n_f+\frac{32}{243}n_g\log{\frac{\mu^2}{\boldsymbol{q}^2}}-\frac{160}{729}n_g+\frac{4}{9}\frac{\tilde{Y}}{\sin{\theta_W}^2}\log{\frac{m_H^2}{\mu^2}}.
\end{align}
\end{subequations}
The factor, $n_f$ defines the number of light quark flavours, $n_g$ is the number of lepton generations, $\tilde{Y}\equiv Y/\alpha$ and $\theta_W$ is the weak mixing angle. The leading $\mathcal{O}(\alpha_s\alpha)$ correction has a colour factor which is correct in the colour singlet configuration, for the colour octet result one replaces each factor of $C_F$ with minus $\frac{1}{2C_A}$.

We note that taking the large vacuum expectation value limit as is done here can be further applied to BSM theories with Higgs-like fields and higher symmetry breaking scales. On the other hand, if we choose to include all interactions of the SM consistently in the static limit, $|\boldsymbol{q}|\ll m_Q, m_{W,Z,H}$, then the more computationally intensive scattering amplitude approach seems to be the safest path; we look at this next.
\section{Scattering Amplitude Approach}
\label{sec:ScattAmp}
Although we have derived the static potential in the Wilson loop approach, it is also worth cross-checking this with the original, scattering amplitude approach \cite{lifshitz1965quantum}. In this way, one can compute the potential directly in momentum space from the on-shell quark-anti-quark scattering amplitude in the static or small momentum transfer limit. Although this is a textbook result in QED \cite{berestetskii1982quantum}, we calculate here in QCD to verify that it matches the Wilson loop result.  The idea of the calculation is to study the QCD scattering amplitude of the process: 
\begin{equation}
    Q(p)+\bar{Q}'(p')\rightarrow Q(p-q)+\bar{Q}'(p'+q),
\end{equation} 
where $q=(0,|\boldsymbol{q}|)$, in the limit of non-relativistic scattering, $m_Q\gg|\boldsymbol{q}|$. There are various ways to parametrise the momenta, we choose to minimise the algebra by employing light-cone coordinates. Taking the initial particles to be moving along the $z$-axis, 
and introducing two light-cone vectors, $n_{\pm}^\mu=(1,0,0,\mp 1)^\mu$, then any momentum is expressible as,
\begin{equation}
k^\mu = \frac{1}{2}\left( n^\mu_- k_+ + n^\mu_+ k_- \right) + {k}^\mu_T,
\end{equation}
where $k^{\mu}_T$ represents the remaining transverse components. This choice of coordinates leads to useful identities,
\begin{equation}
   n++\cdot n_-=2,\quad n_{\pm}^2=0,\quad n_{\pm}\cdot k_T=0,\quad k_{\pm}=k_{\pm}\cdot n_{\pm}=k_0\pm k_3,
\end{equation}
and the scalar product can be re-written as,
\begin{equation}
k\cdot q=\frac{1}{2}(k_+\cdot q_-+k_-\cdot q_+)-{\boldsymbol{k}}_T\cdot{\boldsymbol{q}}_T\quad \Rightarrow \quad k^2=k_+\cdot k_--{\boldsymbol k}_T^2.
\end{equation}
In our case we have four-vectors, $p$ and $p'$, which satisfy,
\begin{equation}
p_+ p_-=m_Q^2,\quad p'_{\pm}=p_\mp,\quad p_{\pm}=\sqrt{m_Q^2+p^2}\pm|{\boldsymbol{ p}}|.
\end{equation}
Assuming we know the transverse part of the transfer momentum, ${\boldsymbol q}_T$, we may fix $q^+$ and $q^-$ in such a way that final particle momenta are on-shell,
\begin{equation}
\left\{ \begin{array}{c}
(p-q)^2=m_Q^2 \\
(p'+q)^2=m_Q^2
\end{array} \right. \Rightarrow \left\{ \begin{array}{c}
(p_+-q_+)(p_--q_-)-{\boldsymbol q}_T^2=m_Q^2 \\
(p_-+q_+)(p_++q_-)-{\boldsymbol q}_T^2=m_Q^2
\end{array} \right.
\end{equation}
Solving this system of equations and substituting the explicit expressions for $p_{\pm}$, we obtain,
\begin{align}
&q_+=-q_-={\cal P}-\bar{\cal P},  \label{eq:sol-qpm}
\end{align} 
such that,
\begin{align}
   & {\cal P}=|{\boldsymbol p}|,\quad \bar{\cal P}=\sqrt{{\boldsymbol p}^2-{\boldsymbol q}_T^2},
\end{align}
 and therefore, ${\boldsymbol q}_T^2={\cal P}^2-\bar{\cal P}^2$, meaning we may express everything in terms of ${\cal P}$ and $\bar{\cal P}$. As these parameters are independent of $m_Q$, we infer that ${\cal P}$, $\bar{\cal P}\ll m_Q$, allowing a safe taking of the leading power dependence of the amplitude at $m_Q\to\infty$. To proceed we need to express all scalar products in terms of our new parameters,
\begin{eqnarray*}
q^2 &=& q_+\cdot q_--{\boldsymbol q}_T^2=2{\cal P}(\bar{\cal P}-{\cal P}),\\
p\cdot q  &=& \frac{1}{2}(p_+\cdot q_- + p_-\cdot q_+)={\cal P}(\bar{\cal P}-{\cal P}), \\
p'\cdot q  &=& \frac{1}{2}(p_-\cdot q_- + p_+\cdot q_+)=-{\cal P}(\bar{\cal P}-{\cal P}), \\
p\cdot p' &=& \frac{1}{2}(p_+^2+p_-^2)= m_Q^2+2{\cal P}^2. 
\end{eqnarray*}
With this set of coordinate re-definitions we may now proceed and calculate the static potential in QCD.
\subsection{QCD}
\label{ssec:SAQCD}
In the pure QCD case, the one-loop bare amplitude will be proportional to the Born amplitude and will be UV and IR finite. The expression for scattering amplitude in perturbation theory up to NLO in the Fourier-transformed potential, $U({\boldsymbol q})$, reads \cite{berestetskii1982quantum},
\begin{equation}
f({\boldsymbol k},{\boldsymbol k}')=-\frac{m_*}{2\pi\hbar^2} \left[U({\boldsymbol k}-{\boldsymbol k}')+\frac{2m_*}{\hbar^2} \int \frac{d^3{\boldsymbol l}}{(2\pi)^3} \frac{U({\boldsymbol k}'-{\boldsymbol l}) U({\boldsymbol l}-{\boldsymbol k})}{{\boldsymbol k}^2-{\boldsymbol l}^2+i0}+O(U^3) \right], 
\end{equation}
where ${\boldsymbol k}={\boldsymbol p}$, ${\boldsymbol k}'={\boldsymbol p}-{\boldsymbol q}$ and $m_*=m_Q/2$ is the effective mass of the scattering particles. For the Coulomb potential, the integral is IR-divergent but we can calculate this in dimensional regularization. The terms we obtain from this procedure should match the corresponding terms in the Wilson loop approach. The UV-divergences will be removed by renormalization of coupling and mass in the $\overline{\text{MS}}$ scheme, while IR divergence in the NR limit is known to come exclusively from the long-range Coulomb interaction and will be removed by on-shell wave-function renormalisation (WFR). 


With our scattering amplitudes, represented in figure \ref{fig:SP1} with strong interactions alone, we take the NR limit of the calculated amplitudes. More specifically, we expand the Dirac spinor chains in terms of Pauli matrices and Pauli spinors, taking the $\boldsymbol{q}^2\rightarrow 0$ limit. Next, we pick out the terms of $\mathcal{O}(1/\boldsymbol{q}^2)$ and sandwiched by Pauli spinors alone, dropping terms with insertions of Pauli matrices, i.e. spin-dependent terms, as these contribute at $\mathcal{O}(\boldsymbol{q}^2/m_Q^2)$. This leads us to to the following renormalised colour singlet potential in the $\overline{\text{MS}}$ scheme,
\begin{eqnarray}
V(\boldsymbol{q}) &=& -\frac{4\pi\alpha_s}{\boldsymbol{q}^2} \left\{ 1+\frac{\alpha_s}{4\pi}\left[(-2\pi i) \frac{mC_F}{{\cal P}} \ln\frac{\mu^2}{\boldsymbol{q}^2}+ \beta_0\ln\frac{\mu^2}{\boldsymbol{q}^2} + \frac{31}{9}C_A-\frac{20}{9}n_f\right] \right\},
\end{eqnarray}
where $\beta_0=11-2n_f/3$ is the first coefficient of the QCD $\beta$-function and the imaginary term proportional to $(-2\pi i)$ is the so-called Coulomb contribution which is known to appear \cite{lifshitz1965quantum}. The real part is exactly the QCD static potential at one-loop order which is identical to the result one obtains with the Wilson loop approach, as required.

\subsection{Standard Model}
\label{ssec:SASM}
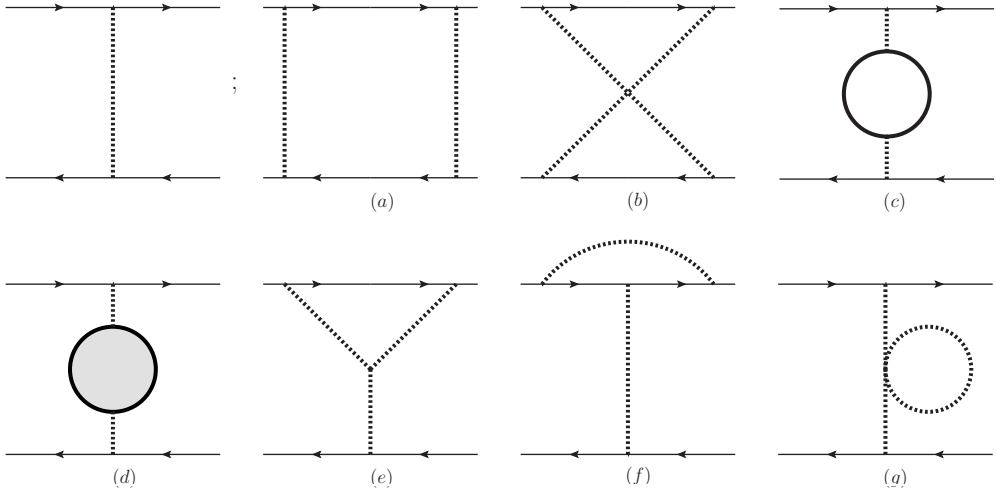
\begin{figure}[bp]
\centering
\begin{center}
\scalebox{0.5}{
\fcolorbox{white}{white}{
  \begin{picture}(739,359) (31,-29)
    \SetWidth{1.0}
    \SetColor{Black}
    \Line[arrow,arrowpos=0.5,arrowlength=5,arrowwidth=2,arrowinset=0.2](32,327)(112,327)
    \Line[arrow,arrowpos=0.5,arrowlength=5,arrowwidth=2,arrowinset=0.2](112,327)(192,327)
    \Line[arrow,arrowpos=0.5,arrowlength=5,arrowwidth=2,arrowinset=0.2](192,199)(112,199)
    \Line[arrow,arrowpos=0.5,arrowlength=5,arrowwidth=2,arrowinset=0.2](112,199)(32,199)
    \Line[arrow,arrowpos=0.5,arrowlength=5,arrowwidth=2,arrowinset=0.2](224,327)(304,327)
    \Line[arrow,arrowpos=0.5,arrowlength=5,arrowwidth=2,arrowinset=0.2](304,327)(384,327)
    \Line[arrow,arrowpos=0.5,arrowlength=5,arrowwidth=2,arrowinset=0.2](304,199)(224,199)
    \Line[arrow,arrowpos=0.5,arrowlength=5,arrowwidth=2,arrowinset=0.2](384,199)(304,199)
    \Line[arrow,arrowpos=0.5,arrowlength=5,arrowwidth=2,arrowinset=0.2](416,327)(496,327)
    \Line[arrow,arrowpos=0.5,arrowlength=5,arrowwidth=2,arrowinset=0.2](496,327)(576,327)
    \Line[arrow,arrowpos=0.5,arrowlength=5,arrowwidth=2,arrowinset=0.2](496,199)(416,199)
    \Line[arrow,arrowpos=0.5,arrowlength=5,arrowwidth=2,arrowinset=0.2](576,199)(496,199)
    \Line[arrow,arrowpos=0.5,arrowlength=5,arrowwidth=2,arrowinset=0.2](32,119)(112,119)
    \Line[arrow,arrowpos=0.5,arrowlength=5,arrowwidth=2,arrowinset=0.2](112,119)(192,119)
    \Line[arrow,arrowpos=0.5,arrowlength=5,arrowwidth=2,arrowinset=0.2](112,-9)(32,-9)
    \Line[arrow,arrowpos=0.5,arrowlength=5,arrowwidth=2,arrowinset=0.2](192,-9)(112,-9)
    \Line[arrow,arrowpos=0.5,arrowlength=5,arrowwidth=2,arrowinset=0.2](416,119)(496,119)
    \Line[arrow,arrowpos=0.5,arrowlength=5,arrowwidth=2,arrowinset=0.2](496,119)(576,119)
    \Line[arrow,arrowpos=0.5,arrowlength=5,arrowwidth=2,arrowinset=0.2](496,-9)(416,-9)
    \Line[arrow,arrowpos=0.5,arrowlength=5,arrowwidth=2,arrowinset=0.2](576,-9)(496,-9)
    \Line[arrow,arrowpos=0.5,arrowlength=5,arrowwidth=2,arrowinset=0.2](608,119)(688,119)
    \Line[arrow,arrowpos=0.5,arrowlength=5,arrowwidth=2,arrowinset=0.2](688,119)(768,119)
    \Line[arrow,arrowpos=0.5,arrowlength=5,arrowwidth=2,arrowinset=0.2](688,-9)(608,-9)
    \Line[arrow,arrowpos=0.5,arrowlength=5,arrowwidth=2,arrowinset=0.2](768,-9)(688,-9)
    \SetWidth{3.0}
    \Line[dash,dashsize=2](112,327)(112,199)
    \Line[dash,dashsize=2](240,327)(240,199)
    \Line[dash,dashsize=2](368,327)(368,199)
    \Line[dash,dashsize=2](432,327)(560,199)
    \Line[dash,dashsize=2](560,327)(432,199)
    \Arc[dash,dashsize=2,clock](496,71)(80,143.13,36.87)

    \Line[dash,dashsize=2](496,-9)(496,119)
    \Line[dash,dashsize=2](688,119)(688,-9)
    \Arc[dash,dashsize=2](720,55)(32,180,540)
    \GOval(112,55)(32,32)(0){0.882}
    \Line[dash,dashsize=2](112,119)(112,87)
    \Line[dash,dashsize=2](112,23)(112,-9)
    \Text(201,263)[lb]{\huge{\Black{$;$}}}
    \Text(305,174)[lb]{\Large{\Black{$(a)$}}}
    \Text(497,175)[lb]{\Large{\Black{$(b)$}}}
    \Text(689,174)[lb]{\Large{\Black{$(c)$}}}
    \Text(112,-34)[lb]{\Large{\Black{$(d)$}}}
    \Text(305,-34)[lb]{\Large{\Black{$(e)$}}}
    \Text(496,-33)[lb]{\Large{\Black{$(f)$}}}
    \Text(689,-34)[lb]{\Large{\Black{$(g)$}}}
    \SetWidth{1.0}
    \Line[arrow,arrowpos=0.5,arrowlength=5,arrowwidth=2,arrowinset=0.2](609,326)(689,326)
    \Line[arrow,arrowpos=0.5,arrowlength=5,arrowwidth=2,arrowinset=0.2](689,326)(769,326)
    \Line[arrow,arrowpos=0.5,arrowlength=5,arrowwidth=2,arrowinset=0.2](689,198)(609,198)
    \Line[arrow,arrowpos=0.5,arrowlength=5,arrowwidth=2,arrowinset=0.2](769,198)(689,198)
    \SetWidth{3.0}
    \Arc(689,262)(32,180,540)
    \Line[dash,dashsize=2](689,326)(689,294)
    \Line[dash,dashsize=2](689,230)(689,198)
    \SetWidth{1.0}
    \Line[arrow,arrowpos=0.5,arrowlength=5,arrowwidth=2,arrowinset=0.2](224,119)(304,119)
    \Line[arrow,arrowpos=0.5,arrowlength=5,arrowwidth=2,arrowinset=0.2](304,119)(384,119)
    \Line[arrow,arrowpos=0.5,arrowlength=5,arrowwidth=2,arrowinset=0.2](304,-9)(224,-9)
    \Line[arrow,arrowpos=0.5,arrowlength=5,arrowwidth=2,arrowinset=0.2](384,-9)(304,-9)
    \SetWidth{3.0}
    \Line[dash,dashsize=2](240,119)(304,55)
    \Line[dash,dashsize=2](304,55)(368,119)
    \Line[dash,dashsize=2](304,55)(304,-9)
  \end{picture}}
}
\end{center}
\caption{Diagrams that contribute to the static potential to one-loop order where the dotted lines represent possible bosonic propagators. The shaded and hollow circles represent light fermion/ghost and bosonic self-energies, respectively. }
\label{fig:SP1}
\end{figure}

We may now extend this approach in QED/QCD to the standard model, expanding the scattering amplitudes illustrated in figure \ref{fig:SP1} in the non-relativistic limit, and taking the real part of the expression to be the static potential. We will present analytical expressions in reasonable limits and leave a numerical comparison of the full expressions to section \ref{scn:disc}.
As in the QCD case, we investigate the following process at one-loop,
\begin{equation}
    Q_1(p)+\bar{Q}_2(p')\rightarrow Q_1(p-q)+\bar{Q}_2(p'+q),
\end{equation} 
where $q=(0,|\boldsymbol{q}|)$ is the transfer momentum. The limit of NR scattering in the SM is given by, $|\boldsymbol{q}|\ll m_{1,2}, m_{W,Z,H}$, where the subindices 1,2 denotes the possibility of working with different particles (different masses). Therefore, we have three cases to consider, 
\begin{equation}
    V_{ij}^{SM}=V^{QCD}+V^{QED}+\delta V^{SM}_{ij}
\end{equation}
where $\delta V^{SM}_{(i,j)}$ is the one-loop correction from contributions outside of pure QCD, the leading of which will be of $\mathcal{O}(\alpha\alpha_s)$. We note that flavour changing is permitted in the SM, so we take the internal quark masses to be non-zero and maintain consistency. To present our large expression for $\delta V_{ij}^{SM}$ concisely, we consider the limit,  $m_1\gg m_{W,Z,H} \gg m_2 \gg \boldsymbol{q}^2$, which is valid for top/bottom quarks, i.e. $m_{1,2}\leftrightarrow m_{t,b}$; other limits including the more physical limit, $m_1\sim m_{W,Z,H} \gg m_2 \gg \boldsymbol{q}^2$, can be considered from the full expressions attached in an ancillary file. In the regime examined we have the following leading terms,
\begin{align}
    \delta V_{(1,1)}^{SM}&=\frac{C_F\alpha\alpha_s}{4\pi {\boldsymbol{q}^2}}\left\lbrace \frac{5  c_w^2}{16 s_w^2 }+\frac{467  c_w^2}{144}-\frac{5 }{16
   s_w^2}-\frac{32 }{9} +\left[ \frac{3 c_w^2}{16 s_w^2}-\frac{137  c_w^2}{48 } \right.\right. \nonumber \\ & \left.\left.-\frac{3 }{16
   s_w^2}+\frac{8 }{3} \right]\log{\frac{m_1^2}{m_Z^2}}-\left[ \frac{3  c_w^2}{16 s_w^2}-\frac{155  c_w^2}{144}-\frac{3 }{16
   s_w^2}+\frac{8 }{9}\right]\log{\frac{\mu^2}{m_Z^2}} \right\rbrace \nonumber \\ &
   +\frac{\alpha^2}{4\pi\boldsymbol{q}^2}\left\lbrace \frac{400 n_f}{729}+\frac{5 c_w^2}{36 s_w^2}+\frac{467 c_w^2}{324}+\frac{80
   n_g}{81}-\frac{5}{36 s_w^2}-\frac{152}{81}  \right. \nonumber \\ & \left. +\left[ \frac{c_w^2}{12 s_w^2}-\frac{137 s_w^2}{108}-\frac{1}{12 s_w^2}+\frac{32}{27}\right]\log{\frac{m_1^2}{m_Z^2}}-\left[ \frac{c_w^2}{12 s_w^2}-\frac{155 c_w^2}{324}-\frac{1}{12 s_w^2}+\frac{32}{81}\right]\log{\frac{\mu^2}{m_Z^2}} \right. \nonumber \\ & \left. -\left[ -\frac{2 c_w^2}{3 s_w^2}+\frac{16 c_w^2}{9}+\frac{2}{3 s_w^2}+\frac{2}{3}\right]\log{\frac{\mu^2}{m_W^2}}+\left[\frac{80 n_f}{243}+\frac{16 n_g}{27}\right]\log{\frac{\mu^2}{\boldsymbol{q}^2}} \right\rbrace \\  
   &\equiv\frac{\alpha\alpha_s}{4\pi\boldsymbol{q}^2}c_{(1,1)}
   +\frac{\alpha^2}{4\pi\boldsymbol{q}^2}\left\lbrace d_{(1,1)}+e_{(1,1)}\log{\frac{\mu^2}{\boldsymbol{q}^2}} \right\rbrace,
\end{align}
such that $c_w\equiv \cos{\theta_W}=m_W/m_Z$ and $s_W\equiv\sin{\theta_W}$ where $\theta_W$ is the weak mixing angle. In the unequal mass case, $ \delta V_{(1,2)}^{SM}= \delta V_{(2,1)}^{SM}$, we have the following result,
\begin{align}
    \delta V_{(1,2)}^{SM}&=\frac{C_F\alpha\alpha_s}{4\pi\boldsymbol{q}^2}\left\lbrace \frac{10}{9}-\frac{21}{16s_w^2}+\frac{27}{16c_w^2}-\frac{3\pi y_{w_1}}{8s_w^2 }+\frac{1}{3}\log{\frac{m_2^2}{\mu^2}}+\frac{13}{9}\log{\frac{m_1^2}{m_Z^2}}\right. \nonumber \\ & \left. +\left[ \frac{1}{4}\log{\frac{m_1^2}{m_W^2}}-\frac{1}{16}\log{\frac{m_2^2}{m_Z^2}}\right]\frac{1}{s_w^2}-\left[ \frac{4}{3}\log{\frac{m_1^2}{m_Z^2}}+\frac{10}{18}\log{\frac{m_2^2}{\mu^2}}+\frac{13}{72}\log{\frac{m_2^2}{m_Z^2}}\right]\frac{1}{c_w^2} \right\rbrace \nonumber \\ &
   +\frac{\alpha^2}{4\pi\boldsymbol{q}^2}\left\lbrace \frac{52}{81}-\frac{40}{81}n_g+\frac{8}{27}n_g\log{\frac{\mu^2}{\boldsymbol{q}^2}}-\frac{200}{729}n_f+\frac{40}{243}n_f\log{\frac{\mu^2}{\boldsymbol{q}^2}} +\frac{14}{9}\log{\frac{\mu^2}{m_W^2}}\right.\nonumber \\ & \left.+\frac{14}{81}\log{\frac{\mu^2}{m_2^2}}+\frac{4}{81}\log{\frac{m_Z^2}{m_2^2}}+\frac{22}{81}\log{\frac{m_Z^2}{m_1^2}}   +\left[\frac{7}{24}-\frac{\pi y_{w_1}}{12}+\frac{1}{72}\log{\frac{m_2^2}{m_Z^2}}+\frac{1}{18}\log{\frac{m_1^2}{m_W^2}} \right]\frac{1}{s_w^2}\right. \nonumber \\ & \left. - \left[ \frac{3}{8}+\frac{10}{81}\log{\frac{\mu^2}{m_2^2}}+\frac{13}{648}\log{\frac{m_Z^2}{m_2^2}}+\frac{8}{27}\log{\frac{m_Z^2}{m_1^2}}\right]\frac{1}{c_w^2}\right\rbrace \\  &\equiv\frac{\alpha\alpha_s}{4\pi\boldsymbol{q}^2}c_{(1,2)}
   +\frac{\alpha^2}{4\pi\boldsymbol{q}^2}\left\lbrace d_{(1,2)}+e_{(1,2)}\log{\frac{\mu^2}{\boldsymbol{q}^2}} \right\rbrace,
\end{align}
such that $y_{w_i}  \equiv \frac{m_i}{m_W}$. Lastly, for $\delta V_{22}^{SM}$, we take the further approximation of $m_{W,Z,H}\sim M_{ew}$, reducing the expression size further,
\begin{align}
     \delta V_{(2,2)}^{SM}&=\frac{C_F\alpha\alpha_s}{4\pi\boldsymbol{q}^2}\left\lbrace \frac{8}{6}\log{\frac{m_2^2}{M_{ew}^2}}+\frac{1}{72}\log{\frac{M_{ew}^2}{\mu^2}} -\frac{83}{36}\right\rbrace \frac{\alpha^2}{4\pi\boldsymbol{q}^2}\left\lbrace \frac{100}{729}n_f+\frac{20}{81}n_g-\frac{67}{324} \right. \nonumber \\ & \left. +\left[\frac{20}{243}n_f+\frac{4}{27}n_g\right]\log{\frac{\mu^2}{\boldsymbol{q}^2}}+\frac{151}{324}\log{\frac{M_{ew}^2}{\mu^2}}+\frac{4}{54}\log{\frac{m_2^2}{\mu^2}} \right\rbrace 
     \\  &\equiv\frac{\alpha\alpha_s}{4\pi\boldsymbol{q}^2}c_{(2,2)}
   +\frac{\alpha^2}{4\pi\boldsymbol{q}^2}\left\lbrace d_{(2,2)}+e_{(2,2)}\log{\frac{\mu^2}{\boldsymbol{q}^2}} \right\rbrace.
\end{align}
Note that the above are color singlet results, again the simple replacement $C_F\mapsto -1/(2C_A)$ produces the colour octet results for the $\mathcal{O}(\alpha\alpha_s)$ contributions.
Thus, we have now fully expressed the SM static potential to one-loop order, which was not possible in the Wilson loop approach.
\section{Applications}
\label{sec:EFTandMD}
The potential on its own represents a fundamental concept, which provides not only potential models that have been astonishingly successful in the description of quarkonia, but also a deeper understanding of confinement. From a more phenomenological standpoint however, the primary interest lies in heavy quark pair production at threshold \cite{brambilla2000potential,beneke2015next}. The static potential appears in heavy quark effective theories, for instance in pNRQCD \cite{pineda1997effective,brambilla2000potential}, where both the singlet and octet potential appears as Wilson coefficients of the theory. Moreover, there has been significant interest in producing high precision quark mass definitions appropriate for processes occurring at threshold, the most popular of which are the 1S and potential subtracted (PS) masses \cite{beneke1998quark,hoang1999b}. We will summarise these two applications and comment on the effect of incorporating the SM potential,  or EW corrections to the QCD potential, to these results.
\subsection{Potential NRQCD}
\label{ssec:SPpnrqcd}
The effective theory, pNRQCD, is an often employed extension of NRQCD; the difference between these two theories is that pNRQCD takes further advantage of the hierarchy of scales that appear in a particular process. The hierarchy under consideration is taken to be, $m_Q\gg |\boldsymbol{p}|\sim m_Qv\gg E\sim m_Qv^2$, where pNRQCD takes into account the ultrasoft (US) scale, $E\sim m_Qv^2$, which is neglected in NRQCD \cite{pineda1998effective}. For one to take into account the US scale, one is implored to alter the Lagrangian of NRQCD by including the following terms,
\begin{equation}
    \mathcal{L}_{pNRQCD}=\mathcal{L}_{NRQCD}^{US}+\mathcal{L}_{pot},
\end{equation}
where $\mathcal{L}_{NRQCD}^{US}$ is identical to $\mathcal{L}_{NRQCD}$ with all gluons taken to be in the US regime. The second term, $\mathcal{L}_{pot}$, is of particular interest to us and it arises from the Schrodinger equation,
\begin{equation}
    \mathcal{L}_{pot}=-\int d^3\boldsymbol{x}_1d^3\boldsymbol{x}_2\psi^{\dagger}(t,\boldsymbol{x}_1)\chi(t,\boldsymbol{x}_2)V(\boldsymbol{r})\chi^{\dagger}(t,\boldsymbol{x}_2)\psi(t,\boldsymbol{x}_1),
\end{equation}
such that $\boldsymbol{p}_j=-i\boldsymbol{\nabla}_j$ and $\boldsymbol{S}_j=\boldsymbol{\sigma}_j/2$,
where $j=1,2$ act on the fermion and antifermion, respectively. Moreover, the fermion and antifermion spin indices are contracted with the indices of $V(\boldsymbol{r})$, which are not explicitly displayed. The potential in this expression, $V(\boldsymbol{r})$, is precisely the QCD static potential. There are implicitly two terms in this Lagrangian, depending on if the wave functions are color singlet or octet with corresponding singlet and octet potentials, respectively. By inspection, the potential $V_{s,o}(\boldsymbol{r})$ contains both the expansion parameter and Wilson coefficients of this effective theory. Conversely, this effective can be seen as defining the static potential, i.e. any term matching to the EFT is the static potential. Whence, when employing this effective theory one should include the EW corrections to the QCD static potential at NLO as they are comparative to NNLO QCD corrections.

\subsection{Threshold Masses}
It is well known that contrary to intuition, the notion of a quark pole mass, is, in fact, inadequate for accurate calculations
of heavy quark cross-sections near threshold. The loss of accuracy is due to the existence of IR renormalons that have been studied in various contexts \cite{beneke1999renormalons}.
The PS and 1S masses are by far the most used threshold mass definitions that evade the renormalon problem by employing the static potential. The PS mass is slightly more involved phenomenologically as it introduces a new factorisation scale, $\mu_F$, in it's definition. Before introducing these mass definitions, we must first examine the Fourier transform of our potential.
\subsubsection{Fourier Transform}
We are now able to compute the standard model analogue of the well-known Coulomb potential, i.e. the SM potential in position space. From this, we may obtain the corrections to the PS mass and compare to the pure QCD result. In order to simplify our expressions, it is convenient to introduce the notation \cite{peter1997static},
\begin{equation}
    \mathcal{F}(\boldsymbol{r},\mu,u)=\mu^{2u}\int\slashed{d}^3\boldsymbol{q}\frac{e^{i\boldsymbol{q}\cdot{\boldsymbol{r}}}}{(\boldsymbol{q}^2)^{1+u}},
\end{equation}
for the Fourier transform of a general power of $1/\boldsymbol{q}^2$. We then employ a Schwinger parameter,
\begin{equation}
    \frac{1}{(\boldsymbol{q}^2)^{1+u}}=\frac{1}{\Gamma[1+u]}\int_0^{\infty}dx x^u e^{-x\boldsymbol{q}^2}.
\end{equation}
There are various representations of $\mathcal{F}$, the ones which are useful to us are,
\begin{align}
    \mathcal{F}(|\boldsymbol{r}|,\mu,u)&= \frac{(\mu |\boldsymbol{r}|)^{2u}}{4\pi^2|\boldsymbol{r}|}\frac{\Gamma[\frac{1}{2}+u]\Gamma[\frac{1}{2}-u]}{\Gamma[1+2u]} \\ & = \frac{(\mu |\boldsymbol{r}| e^{\gamma_E})^{2u}}{4\pi |\boldsymbol{r}|}\exp{\left( \sum_{n=2}^{\infty}\frac{\zeta(n)u^n}{n}(2^n-1-(-1)^n) \right)},
\end{align}
where the first and second formulas are applicable if $-1<u<1/2$ and $|u|<1/2$, respectively. By inspection of the static potential, we need the Fourier transform of $\log^m{(\mu^2/\boldsymbol{q}^2)}$, which is easily attainable from $\mathcal{F}$ since,
\begin{equation}
    \log^m{\frac{\mu^2}{\boldsymbol{q}^2}}=\left[ \frac{\partial^m}{\partial u^m}\left(\frac{\mu^2}{\boldsymbol{q}^2}\right)\right]\biggr\rvert_{u = 0},
\end{equation}
and therefore, 
\begin{equation}
    \int \slashed d^3{q} \log^m\left({\frac{\mu^2}{\boldsymbol{q}^2}}\right)\frac{e^{i{\boldsymbol{q}\cdot \boldsymbol{r}}}}{ {\boldsymbol{q}}^2}=\left(\frac{\partial^m}{\partial u^m}\mathcal{F}(r,\mu,u)\right)\biggr\rvert_{u = 0}.
\end{equation}
Whence the color singlet potential (in the $\overline{\text{MS}}$ scheme) in position space is,
\begin{align}
    V^{SM}_{(i,j)}(\boldsymbol{r})&=-C_F\frac{\alpha_s}{|\boldsymbol{r}|}\left\lbrace 1+\frac{\alpha_s}{4\pi}(2\beta_0\log{\mu r'}+a_1)\right\rbrace \nonumber \\ &-\frac{\alpha}{|\boldsymbol{r}|}\left\lbrace Z_iZ_j+\frac{\alpha_s}{4\pi}c_{(i,j)}+\frac{\alpha}{4\pi}\left(d_{(i,j)}+2e_{(i,j)}\log{\mu r'}\right)\right\rbrace,
\end{align}
where $Z_i$ is the fractional charge of the incoming and outgoing heavy quarks, in our calculation we take $Z_1=+2/3$, $Z_2=-1/3$ and $r'=|\boldsymbol{r}|e^{\gamma_E}$. The result may be considered in several ways because employing it as it
stands is not reasonable due to the the possibly large logarithms. Therefore, selecting a renormalisation scale, $\mu$, that reduces the higher order corrections would be ideal. The first choice and the one most often employed is $\mu_1=1/|\boldsymbol{r}|$ or $\mu_1=1/r'$; other choices include selecting $\mu$ in such a way the first order pure QCD coefficient is removed entirely or solely removing all $n_f$-dependency from the coefficients \cite{brodsky1983elimination}. With the Fourier transformed potential at hand, we may now consider the PS mass.
\subsubsection{Potential Subtracted Mass}
\label{ssec:SPPS}
It is well-known that the coordinate space potential is more sensitive to long distances than the potential in momentum space and its leading power correction is linear in $\Lambda_{QCD}|\boldsymbol{r}|$ \cite{beneke1998quark}. The implication is that the expansion of the QCD coordinate-space potential in $\alpha_s(e^{-\gamma_E}|\boldsymbol{r}|)$ diverges as,
\begin{equation}
    \sum_nr_n\alpha_s(e^{-\gamma_E}|\boldsymbol{r}|)^{n+1}\sim \sum_n(-2\beta_0)^nn!n^b\alpha_s(e^{-\gamma_E}/|\boldsymbol{r}|)^{n+1},
\end{equation}
which is much faster than the expansion of the potential in momentum space. This divergent behaviour has been studied in previous works \cite{aglietti1995renormalons,jezabek1998relation}. It is clear that the rapid divergence originates only from the Fourier transform to coordinate space and is not present in momentum space. Knowing this, one can subtract the leading long-distance contribution and the LO divergent behaviour completely by restricting the Fourier integral $|\boldsymbol{q}|>\mu_f$ with $\mu_f$ a new factorisation scale, which is viewed as an IR regulator. The result is called the `subtracted potential', $V(\boldsymbol{r},\mu_f)$. The subtraction terms can be evaluated order by order in the coupling given $V(\boldsymbol{q})$ to that order. More precisely,
\begin{equation}
    V(\boldsymbol{r},\mu_f)=V(\boldsymbol{r})+2\delta m(\mu_f),
\end{equation}
where,
\begin{equation}
    \delta m(\mu_f)=-\frac{1}{2}\int_{|\boldsymbol{q}|<\mu_f}\slashed{d}^3{\boldsymbol{q}}V(\boldsymbol{q}).
    \label{eqn:dmps}
\end{equation}
To subtract the leading long-distance contribution of order $\Lambda_{QCD}$, it is reasonable to replace the factor, $e^{i\boldsymbol{q}\cdot \boldsymbol{r}}$, by unity in the Fourier transform and this is used as the definition for the subtraction term in the mass definition,
\begin{equation}
   m_{PS}(\mu,\mu_f)=m_{pole}-\delta m(\mu,\mu_f).
\end{equation}
Of course, in this procedure one has only swept the large loop corrections from $\delta m (\mu_f)$ to $m_{PS}(\mu_f)$. However, when $m_{pole}$ is expressed in terms of a short-distance mass parameter such as the $\overline{\text{MS}}$ mass through a perturbative series, this series will also contain large loop corrections \cite{bigi1994pole}. Conveniently, these perturbative corrections cancel with large perturbative corrections to the pole mass in $\delta m (\mu_f)$. In this way, one may determine the  $\overline{\text{MS}}$ mass from threshold cross-sections with better accuracy than the pole mass, the use of which implicitly contains the non-subtracted potential.  Let us now compute $\delta m(\mu_f)$ from the definition in Eq. \eqref{eqn:dmps}, one obtains,
\begin{align}
        \delta m^{SM}_{(i,i)}(\mu_f)&=C_F\frac{\alpha_s}{\pi}\mu_f\left\lbrace 1+\frac{\alpha_s}{4\pi}\left(a_1+2\beta_0\left(\log{\frac{\mu}{\mu_f}}+1\right)\right)\right\rbrace \nonumber \\ &+\frac{\alpha}{\pi}\mu_f\left\lbrace Z_i^2+\frac{\alpha_s}{4\pi}c_{(i,i)}+\frac{\alpha}{4\pi}\left(d_{(i,i)}+2e_{(i,i)}\left(\log{\frac{\mu}{\mu_f}}+1\right)\right)\right\rbrace,
\end{align}
where in the case $i=1$, in the context of the previous limits, one may take this to be the case of the top quark mass, conversely if $i=2$, the bottom quark. For completeness, we avoid taking limits and employ the complete result to get a numerical estimate of the one-loop standard model PS mass. One usually picks $\mu_f=3$ GeV, a typical scale for heavy quarks, and $\mu=m_Z$ to avoid large logarithms in the coefficients, $\lbrace c_{(i,j)},d_{(i,j)}\rbrace$. Moreover, we choose $m_1\rightarrow m_t(m_Z)$, $m_2\rightarrow m_b(m_Z)$, the number of light quarks and lepton generations, $n_f\equiv 2n_u-n_d=2$ and $n_g=3$, respectively (the rest of the parameters are taken from the latest PDG review \cite{tanabashi2018review}). From this we obtain,
\begin{align}
        &\delta m^{SM}_{(1,1)}(\mu_f)=C_F\frac{\alpha_s}{\pi}\mu_f\left\lbrace 1+79.5\frac{\alpha_s}{4\pi}\right\rbrace +\frac{\alpha}{\pi}\mu_f\left\lbrace 0.44-24.1\frac{\alpha_s}{4\pi}+17.7\frac{\alpha}{4\pi}\right\rbrace, \\ &
         \delta m^{SM}_{(2,2)}(\mu_f)=C_F\frac{\alpha_s}{\pi}\mu_f\left\lbrace 1+79.5\frac{\alpha_s}{4\pi}\right\rbrace +\frac{\alpha}{\pi}\mu_f\left\lbrace 0.11-0.45\frac{\alpha_s}{4\pi}+5.1\frac{\alpha}{4\pi}\right\rbrace.
\end{align}
One can clearly note that although, at NLO, the contributions outside of pure QCD/QED are significantly smaller, they still have an impact in high precision calculations and are comparative to NNLO higher order QCD corrections and thus must be taken into account in this mass definition.
\subsubsection{1S Mass}
\label{ssec:SP1S}
The PS mass along with other threshold mass definitions, such as the kinetic and MSR masses \cite{bigi1997high,hoang2008top}, are defined by introducing a new explicit IR factorization scale, $\mu_f$, to remove the IR ambiguity of the pole mass. In contrast, the 1S mass \cite{hoang1999b}, denoted by $m_{1S}$, achieves a similar goal without introducing a new factorization scale. The 1S mass is defined as one-half of the perturbative energy of the 1S heavy $q\bar{q}$ bound state,
\begin{equation}
    m_{1S}(\mu)=\frac{1}{2}(m_{1S}^{q\bar{q}})_{pert}\equiv m_{pole}-\delta m(\mu).
\end{equation}
The ground state energy calculated from the Schr\"{o}dinger equation from elementary quantum mechanics is exactly $(m_{1S}^{q\bar{q}})_{pert}$. At leading order in the small quark velocity expansion (threshold region), the dynamics of a heavy $q\bar{q}$ pair is governed by the Hamiltonian \cite{melnikov1999b},
\begin{equation}
    H=-\frac{\boldsymbol{\nabla}^2}{m_Q}+V(\boldsymbol{r})+U(\boldsymbol{q},\boldsymbol{r}),
\end{equation}
where $m_Q$ is the quark pole mass, $V(\boldsymbol{r})$ is the static potential, the analogue to the Coulomb potential, and $U(\boldsymbol{q},\boldsymbol{r})$ encodes higher order corrections in the small velocity expansion and is the SM analogue of the Breit potential \cite{landau1974course}. The leading contributions at threshold come from the static potential and thus we omit $U(\boldsymbol{q},\boldsymbol{r})$ from our calculation. Solving for the S-wave Green function, we have,
\begin{equation}
    G(E)=\bra{0}\hat{G}(E)\ket{0}=\bra{0}\frac{1}{H-E-i\delta}\ket{0},
\end{equation}
where $\bra{0}$ denotes a position eigenstate with eigenvalue, $|\boldsymbol{r}|=0$, and the Green function has single poles at the exact S-wave energy levels, $E=E_n$,
\begin{equation}
    G(E)\overset{E\rightarrow E_n}{=}\frac{|\psi_n(0)|^2}{E_n-E-i\delta}.
\end{equation}
From this expression, one gets $(m_{1S}^{q\bar{q}})_{pert}=E_1$ and expanding one half $E_1$ in small standard model couplings gives the 1S mass, which is the SM analogue of the well-known Bohr potential from quantum mechanics. We may then find the leading EW corrections at one-loop to the 1S mass with our SM potential, as only the QCD corrections are known, and they have been found to $\text{N}^3\text{LO}$ \cite{ayala2014bottom}. At one-loop up to third order in SM couplings, $\lbrace \alpha,\alpha_s\rbrace$, we have,
\begin{align}
    &\delta m^{SM}_{(i,i)}(r)=\frac{m_i(\alpha Z_i^2+\alpha_sC_F)}{16\pi}\left\lbrace \alpha^2 B_i+\alpha(\alpha_s c_{(i,i)}+2\pi Z_i^2)+\alpha_sC_F(\alpha_sA_i+2\pi) \right\rbrace
\end{align}
such that,
\begin{align}
    & A_i=2\beta_0(l_i+1)+a_1, \\ &
    B_i=2\left(\frac{80}{243}n_f+\frac{16}{27}n_g\right)(l_i+1)+d_{(i,i)},
\end{align}
where $l_i\equiv\log{\frac{\mu}{C_F\alpha_s(\mu)m_i}}$ and $m_i$ is the pole mass. We further note that the IR renormalon cancellation is more subtle in the 1S mass definition, as it is a well-behaved parameter only if the orders of terms in perturbation theory are re-interpreted \cite{hoang1999b}. To see how the leading EW corrections at one-loop alter the 1S mass, we obtain a numerical estimate in a similar fashion to the PS mass and compare the $\mathcal{O}(\alpha\alpha_s,\alpha^2)$ to the $\mathcal{O}(\alpha_s^2)$ terms. We employ the same parameter choices and renormalisation scale as in the PS mass case to obtain the following result,
\begin{align}
        &\delta m^{SM}_{(1,1)}=40.2\alpha_s^2+270.1\alpha_s^3+4.5\alpha^2+17.3\alpha^3+26.8\alpha_s\alpha-25.7\alpha_s^2\alpha+13.3\alpha_s\alpha^2, \\ &
         \delta m^{SM}_{(2,2)}=0.63\alpha_s^2+11.23\alpha_s^3+0.07\alpha^2+0.16\alpha^3+0.42\alpha_s\alpha+3.7\alpha_s^2\alpha+0.48\alpha_s\alpha^2.
\end{align}
It is apparent that at NLO, the contributions outside of pure QCD/QED are significantly smaller, in particular for the case of $m_b$. However, again for high precision calculations it is necessary to include them since, as in the PS mass case, they are comparative to NNLO terms in higher order QCD corrections.
\subsection{Further Applications}
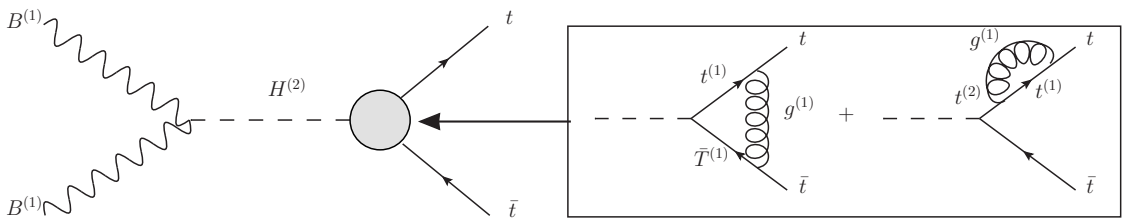
\begin{figure}[H]
    \centering
\begin{center}
    \scalebox{0.56}{
\fcolorbox{white}{white}{
  \begin{picture}(752,152) (45,-27)
    \SetWidth{1.0}
    \SetColor{Black}
    \Photon(65,103)(162,39){7.5}{6}
    \Photon(162,39)(66,-25){7.5}{6}
    \Line[dash,dashsize=10](163,39)(269,39)
    \GOval(289,39)(20,20)(0){0.882}
    \Line[arrow,arrowpos=0.5,arrowlength=5,arrowwidth=2,arrowinset=0.2](303,54)(361,102)
    \Line[arrow,arrowpos=0.5,arrowlength=5,arrowwidth=2,arrowinset=0.2](362,-25)(304,23)
    \Text(40,101)[lb]{\Large{\Black{$B^{(1)}$}}}
    \Text(40,-25)[lb]{\Large{\Black{$B^{(1)}$}}}
    \Text(215,55)[lb]{\Large{\Black{$H^{(2)}$}}}
    \Text(374,104)[lb]{\Large{\Black{$t$}}}
    \Text(376,-26)[lb]{\Large{\Black{$\bar{t}$}}}
    \CBox(414,-26)(782,102){Black}{White}
    \Line[dash,dashsize=10](432,40)(496,40)
    \Line[arrow,arrowpos=0.5,arrowlength=5,arrowwidth=2,arrowinset=0.2](496,40)(560,88)
    \Line[arrow,arrowpos=0.5,arrowlength=5,arrowwidth=2,arrowinset=0.2](560,-8)(496,40)
    \Text(596,38)[lb]{\Large{\Black{$+$}}}
    \Line[dash,dashsize=10](624,40)(688,40)
    \Line[arrow,arrowpos=0.5,arrowlength=5,arrowwidth=2,arrowinset=0.2](688,40)(752,88)
    \Line[arrow,arrowpos=0.5,arrowlength=5,arrowwidth=2,arrowinset=0.2](752,-8)(688,40)
    \Gluon(540,72)(540,7){7.5}{5}
    \GluonArc[clock](719.597,64.984)(19.536,-138.347,-322.042){7.5}{5}
    \Text(560,40)[lb]{\Large{\Black{$g^{(1)}$}}}
    \Text(504,64)[lb]{\Large{\Black{$t^{(1)}$}}}
    \Text(500,8)[lb]{\Large{\Black{$\bar{T}^{(1)}$}}}
    \Text(674,50)[lb]{\Large{\Black{$t^{(2)}$}}}
    \Text(684,86)[lb]{\Large{\Black{$g^{(1)}$}}}
    \Text(728,54)[lb]{\Large{\Black{$t^{(1)}$}}}
    \Text(570,90)[lb]{\Large{\Black{$t$}}}
    \Text(570,-10)[lb]{\Large{\Black{$\bar{t}$}}}
    \Text(762,-10)[lb]{\Large{\Black{$\bar{t}$}}}
    \Text(762,90)[lb]{\Large{\Black{$t$}}}
    \SetWidth{1.4}
    \Line[arrow,arrowpos=1,arrowlength=13,arrowwidth=5.2,arrowinset=0.05](416,38)(322,38)
  \end{picture}}
}
\end{center}
    \caption{Resonant annihilation process of LKP dark matter, $B^{(1)}$, through s-channel $h^{(2)}$}
    \label{fig:lkp}
\end{figure}

As we discussed in the previous section, the number of viable dark matter candidates is rapidly being constrained by precise collider and cosmological experiments, for example, self-interacting theories have been practically ruled out recently by galactic observations \cite{bringmann2017strong}. We may thus focus on computing the static potential of the most viable DM candidates, the lightest Kaluza-Klein particle (LKP) and right-handed neutrinos. As is well understood, Supersymmetry and Extra-Dimensional theories are two strong proponents to an array of issues that belie the standard model \cite{wess1992supersymmetry,appelquist2001bounds}. Dark matter is known to exist but is missing from the SM, both SUSY and Kaluza-Klein theory posit viable dark matter candidates, the properties of which can be understood better in the non-relativistic regime due to their large predicted masses.
A recent static potential calculation for higgsino-wino dark matter found the $SU(2)\times U(1)$ electro-weak static potential between a fermionic triplet in the broken phase of the SM at one-loop order. The NLO terms provided the leading non-relativistic correction to the large resonances (or Sommerfeld effect) in the annihilation cross-section of wino or wino-like dark matter particles. The authors found sizeable modifications from LO of the $\chi_0\chi_0$ annihilation cross-section and determined the shifts of the resonance locations due to the loop correction to the wino potential. Although these results seem promising for future detections, such resonances would also occur in KK theory for the LKP coupling \cite{kakizaki2006relic} to the second excitation of the Higgs as shown in figure \ref{fig:lkp}. 
This will inevitably also be true for right-handed (or sterile) neutrinos due to their possible large mass \cite{davidson2002lower,biondini2013effective,biondini2016cp}, and thus their potential for producing heavy $t\bar{t}$ pairs through the Higgs mechanism in the standard model. Neutrinos are the only matter particles in the Standard Model of particle physics that have been observed with solely left-handed chirality to date. If right-handed neutrinos exist, they could be responsible for several phenomena that have no explanation within the Standard Model, including neutrino oscillations, the baryon asymmetry of the universe, dark matter and dark radiation \cite{palazzo2013phenomenology}. These particles provide us with a test-bed case of our ability to deal with static potentials in theories with spontaneous symmetry breaking as these massive neutrinos solely couple to gravity and the SM Higgs field.

\section{Technical Details}
\label{ssec:TechD}
Our calculations of the one-loop correction to the SM potential employed standard tools. The calculation was performed in general covariant gauge. The diagrams of the type in figure \ref{fig:SP1} were reduced to a few master integrals, which are found analytically since one-loop results are known for all master integrals. For the scattering amplitude approach, we achieved this with the help of \texttt{Mathematica} accompanied by the package, \texttt{FeynCalc} \cite{shtabovenko2016new}, to compute the necessary amplitudes and deal with the algebra. We employed further sub-packages of \texttt{FeynCalc}, such as, \texttt{FeynHelpers} \cite{shtabovenko2017feynhelpers}, which reduces and provides explicit expressions for one-loop scalar integrals by connecting the reduction package, \texttt{fire} \cite{smirnov2020fire6}, with the analytic scalar integrals program, \texttt{Package-X} \cite{patel2015package}. Lastly, we employed the \texttt{FeynOnium} sub-package, which comes equipped with functions for dealing with amplitudes in the NR limit \cite{brambilla2020feynonium}.
For the Wilson loop approach, we employed \texttt{QGRAFS} \cite{nogueira1993automatic} to generate the diagrams and \texttt{FORM} \cite{vermaseren2000new} to deal with the algebra. We also used a \texttt{Mathematica} sub-package, \texttt{LiteRed} \cite{lee2013litered} to reduce our integrals and again \texttt{Package-X} for the analytic one-loop scalar integrals.

\section{Discussion}
\label{scn:disc}
In this paper, we proposed a novel way of studying the static potential for theories that exhibit SSB. We discussed the limitations of the Wilson loop approach for the SM and the need to derive the potential directly from the scattering amplitude. We also mentioned how these techniques could be extended to BSM theories and shed light on which ones would satisfy the criteria to be treated in the Wilson loop fashion. The static potential for the full SM was then presented, and the regimes of applicability considered.  In particular, we showed how our EW corrections to the potential modifies two oft-employed short-distance mass definitions. Moreover, we rounded off each discussion by comparing the size of terms with the SM static potential taken into account versus the QCD potential alone. In doing so, we found the contributions from the EW regime to be significant and comparable to NNLO pure QCD contributions. Therefore, we recommend that the SM potential be employed in future heavy quark high precision studies. Given the framework we now have to build upon, it would be interesting to investigate the static potential of further models, in particular BSM theories with higher symmetry breaking scales, to better understand the non-relativistic regime and explore implications on measurable observables.
\appendix
\acknowledgments
We are indebted to O. Veretin for the many discussions we had on the subject and technical details of the calculation. We would also like to thank M. Nefedov for his insight and guidance in dealing with various aspects of the work.


\bibliographystyle{JHEP}
\bibliography{SPEW.bib}

\end{document}